\documentclass[runningheads]{llncs}
\usepackage{graphicx}

\usepackage{multirow}

\usepackage{multirow}
\usepackage{tabu}
\usepackage{bbm}
\usepackage{diagbox}
\usepackage[english]{babel}
\usepackage{comment}
\usepackage{array}
\usepackage{amstext}
\usepackage{makecell}
\usepackage{caption}
\usepackage{tablefootnote}
\usepackage{babel}
\usepackage{booktabs} 
\usepackage{graphics}

\usepackage[ruled,vlined]{algorithm2e}
\newcommand{\nosemic}{\renewcommand{\@endalgocfline}{\relax}}
\newcommand{\dosemic}{\renewcommand{\@endalgocfline}{\algocf@endline}}
\usepackage{chngcntr}
\usepackage{placeins}
\usepackage{amsfonts}       
\usepackage{nicefrac} 
\usepackage{amsmath,amssymb} 
\usepackage{amstext}

\SetKwInput{KwInput}{Input}                
\SetKwInput{KwOutput}{Output}              
\SetKwInput{KwParameter}{Parameter}              
\usepackage[pagebackref=false,breaklinks=true,letterpaper=true,colorlinks,bookmarks=false]{hyperref}
\newlength\mylen

\newcommand{\ie}{\textit{i.e.}}
\newcommand{\eg}{\textit{e.g.}}

\begin{document}

\title{Asymmetric 3D Context Fusion for Universal Lesion Detection}
\titlerunning{Asymmetric 3D Context Fusion for Universal Lesion Detection}

\author{Jiancheng Yang\inst{1,2,}\thanks{These authors have contributed equally: Jiancheng Yang and Yi He.}  \and Yi He\inst{2,\star} \and Kaiming Kuang\inst{2} \and Zudi Lin\inst{3} \and\\Hanspeter Pfister\inst{3} \and Bingbing Ni\inst{1}\thanks{Corresponding author: Bingbing Ni (nibingbing@sjtu.edu.cn).}}

\authorrunning{J. Yang et al.}

\institute{Shanghai Jiao Tong University, Shanghai, China\\
\email{nibingbing@sjtu.edu.cn}	\\
	\and Dianei Technology, Shanghai, China
	\and Harvard University, Cambridge MA, USA
}

\maketitle              
\begin{abstract}

Modeling 3D context is essential for high-performance 3D medical image analysis. Although 2D networks benefit from large-scale 2D supervised pretraining, it is weak in capturing 3D context. 3D networks are strong in 3D context yet lack supervised pretraining. As an emerging technique, \emph{3D context fusion operator}, which enables conversion from 2D pretrained networks, leverages the advantages of both and has achieved great success. Existing 3D context fusion operators are designed to be spatially symmetric, \ie, performing identical operations on each 2D slice like convolutions. However, these operators are not truly equivariant to translation, especially when only a few 3D slices are used as inputs. In this paper, we propose a novel asymmetric 3D context fusion operator (A3D), which uses different weights to fuse 3D context from different 2D slices. Notably, A3D is NOT translation-equivariant while it significantly outperforms existing symmetric context fusion operators without introducing large computational overhead. We validate the effectiveness of the proposed method by extensive experiments on DeepLesion benchmark, a large-scale public dataset for universal lesion detection from computed tomography (CT). The proposed A3D consistently outperforms symmetric context fusion operators by considerable margins, and establishes a new \emph{state of the art} on DeepLesion. To facilitate open research, our code and model in PyTorch is available at \url{https://github.com/M3DV/AlignShift}.

\end{abstract}

\section{Introduction} \label{sec:intro}

Computer vision for medical image analysis has been dominated by deep learning~\cite{litjens2017survey,shen2017deep}, thanks to the availability of large-scale open datasets~\cite{antonelli2021medical,yang2020medmnist,wei2020mitoem,jin2020deep} and powerful infrastructure. In this study, we focus on 3D medical image analysis, \eg, computed tomography (CT) and magnetic resonance imaging (MRI). Spatial information from 3D voxel grids can be effectively learned by convolutional neural networks (CNNs), while 3D context modeling is still essential for high-performance models. There have been considerable debates over 2D and 3D representation learning on 3D medical images; 2D networks benefit from large-scale 2D pretraining~\cite{deng2009imagenet}, whereas the 2D representation is fundamentally weak in large 3D context. 3D networks learn 3D representations; However, few publicly available 3D medical datasets are large enough for 3D pretraining. 

Recently, there have been a family of techniques that enable building 3D networks with 2D pretraining~\cite{carreira2017quo,qiu2017learning,yang2021reinventing,lin2019tsm,yang2020alignshift}, we refer to it as \emph{3D context fusion operators}. See Sec.~\ref{sec:preliminary} for a review of existing techniques. These operators learn 3D representations while their (partial) learnable weights can be initialized from 2D convolutional kernels. Existing 3D context fusion operators are convolution-like, \ie, either axial convolutions to fuse slice-wise information~\cite{carreira2017quo,qiu2017learning,yang2021reinventing} or shifting adjacent slices~\cite{lin2019tsm,yang2020alignshift}. Therefore, these operators are designed to be spatially \textbf{symmetric}: each 2D slice is operated identically. However, convolution-like operations are not truly translation-equivariant~\cite{luo2016understanding}, due to padding and limited effective receptive fields. In many 3D medical image applications, only a few 2D slices are used as inputs to models due to the memory constraints. It may be meaningless to pursue translation-equivariance in these cases.

In this study, we propose a novel \textbf{asymmetric} 3D context fusion operator (A3D). See Sec.~\ref{sec:a3d} for the methodology details. Basically, given $D$ slices of 3D input features, A3D uses different weights to fuse the input $D$ slices for each output slice. Therefore, the A3D is \textbf{NOT} translation-equivariant. However, it significantly outperforms existing symmetric context fusion operators without introducing large computational overhead in terms of both parameters and FLOPs. We validate the effectiveness of the proposed method by extensive experiments on DeepLesion benchmark~\cite{yan2018deep}, a large-scale public dataset for universal lesion detection from computed tomography (CT). As described in Sec.~\ref{sec:experiments}, the proposed A3D consistently outperforms symmetric context fusion operators by considerable margins, and establishes a new \emph{state of the art} on DeepLesion. 

\section{Methods}

\begin{figure}[tb]
	\includegraphics[width=\textwidth]{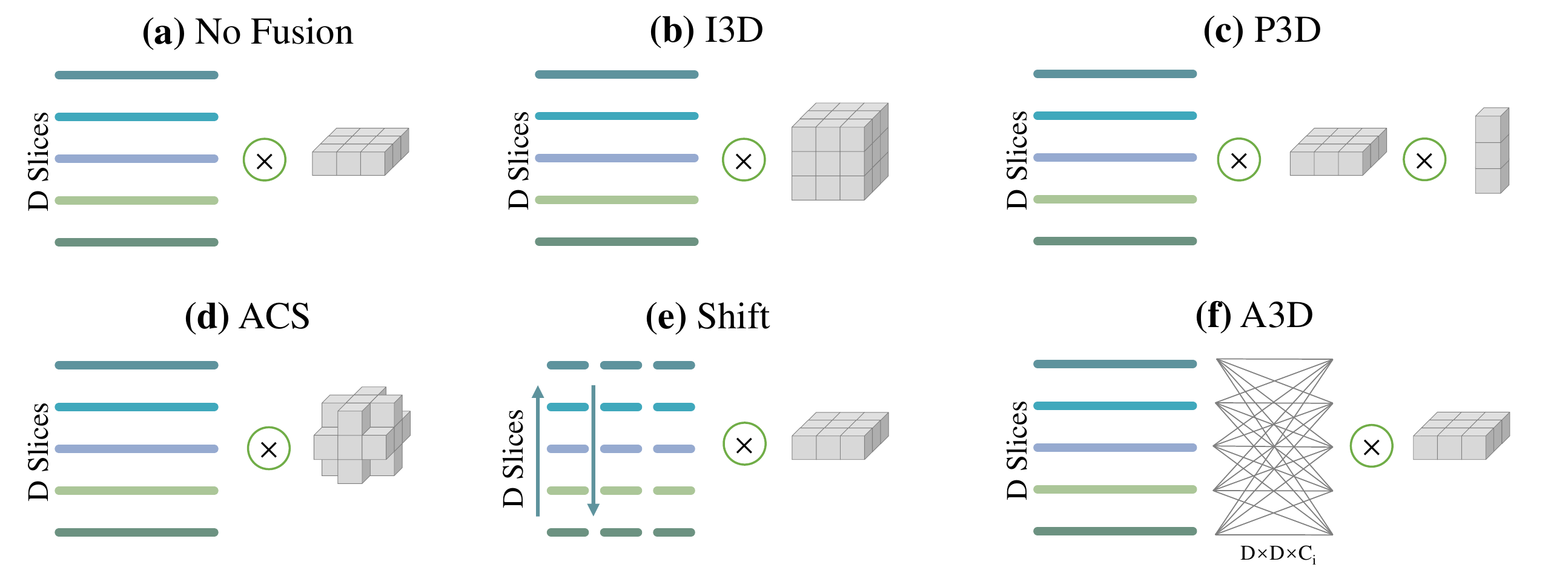}
	\caption{Illustration of various 3D context fusion operators: (a) no fusion, (b) I3D~\cite{carreira2017quo}, (c) P3D~\cite{qiu2017learning}, (d) ACS~\cite{yang2021reinventing}, (e) Shift~\cite{lin2019tsm,yang2020alignshift} and (f) the proposed A3D. In each sub-figure, left: $D$ slices of $C_i$-channel 3D features as inputs; middle: $\otimes$ means convolution; right: illustration of convolutional kernels. } \label{fig:operators}
\end{figure}

\subsection{Preliminary: 3D Context Fusion Operators with 2D Pretraining} \label{sec:preliminary}

In this section, we briefly review the 3D context fusion operators that enable 2D pretraining, including (a) no fusion, (b) I3D~\cite{carreira2017quo}, (c) P3D~\cite{qiu2017learning}, (d) ACS~\cite{yang2021reinventing} and (e) Shift~\cite{lin2019tsm,yang2020alignshift}. As an emerging technique, 3D context fusion operator leverages advantages of both 2D pretraining and 3D context modeling. 

Given a 3D input feature $\boldsymbol{X_i}\in \mathbb{R}^{C_i \times D \times H \times W}$, we would like to obtain a transformed 3D output $\boldsymbol{X_o}\in \mathbb{R}^{C_o \times D \times H \times W}$ with a (pretrained) 2D convolutional kernel $\boldsymbol{W_{2D}}\in \mathbb{R}^{C_i \times C_o \times K \times K}$, where $D\times H \times W$ is the spatial size of 3D features, $C_i$ and $C_o$ are the input and output channels, and $K$ denotes the kernel size. For simplicity, only cases with same padding are considered here. Apart from convolutions, we simply convert 2D pooling and normalization into 3D~\cite{yang2020alignshift}. We then introduce each operator as follows:

\paragraph{(a) no fusion.} We run 2D convolutions on each 2D slice, which is equivalent to 3D convolutions with $\boldsymbol{W_{3D}}\in \mathbb{R}^{C_i \times C_o\times 1 \times K \times K}$ converted from the 2D kernel.

\paragraph{(b) I3D~\cite{carreira2017quo}.} I3D is basically an initialization technique for 3D convolution, $\boldsymbol{W_{I3D}}\in \mathbb{R}^{C_i \times C_o\times K \times K \times K}$ is initialized with $K$ repeats of $\boldsymbol{W_{2D}}/K$, so that the distribution expectation of 3D features is the same as that of 2D features.

\paragraph{(c) P3D~\cite{qiu2017learning}.} P3D convolution is a $1\times K\times K$ 3D convolution followed by a $K\times 1\times 1$ 3D convolution, where the first convolutional kernel is converted from a 2D kernel (same as \textit{no fusion}), and the second is initialized as $[0,...,1,...,0]$ (\eg, $[0,1,0]$ if $K=3$) to make it as no fusion before training.

\paragraph{(d) ACS~\cite{yang2021reinventing}.} ACS runs 2D-like (3D) convolutions in three views of 3D volumes, by splitting the 2D kernel into three 3D kernels: $\boldsymbol{W_{a}}\in \mathbb{R}^{C_i \times C_o^{(a)}\times 1 \times K \times K}$, $\boldsymbol{W_{c}}\in \mathbb{R}^{C_i \times C_o^{(c)}\times 1 \times K \times K}$ and $\boldsymbol{W_{s}}\in \mathbb{R}^{C_i \times C_o^{(s)}\times 1 \times K \times K}$ ($C_o^{(a)}+C_o^{(c)}+C_o^{(s)}=C_o$). 3D context is fused with layer-by-layer ACS transformation without introducing computational cost compared to no fusion.

\paragraph{(e) Shift~\cite{lin2019tsm,yang2020alignshift}.}
Shift is a family of techniques that fuse 3D context by shifting adjacent 2D slices. Take TSM~\cite{lin2019tsm} as an example. It first splits the input feature $\boldsymbol{X_i}\in \mathbb{R}^{C_i \times D \times H \times W}$ by channel into 3 parts: $\boldsymbol{X_i^{+}}\in \mathbb{R}^{C_i^{+} \times D \times H \times W}$,  $\boldsymbol{X_i^{-}}\in \mathbb{R}^{C_i^{-} \times D \times H \times W}$ and $\boldsymbol{X_i^{=}}\in \mathbb{R}^{C_i^{=} \times D \times H \times W}$ ($C_i^{+}+C_i^{-}+C_i^{=}=C_i$).  $\boldsymbol{X_i^{+}}$, $\boldsymbol{X_i^{-}}$ and $\boldsymbol{X_i^{=}}$ are then shifted up, shifted down and kept among the axial axis ($D$ dimension), respectively. Finally, a 3D convolution with $\boldsymbol{W_{3D}}\in \mathbb{R}^{C_i \times C_o\times 1 \times K \times K}$ (as in no fusion) can fuse 3D context with a single slice. AlignShift~\cite{yang2020alignshift} is a shift operator adaptive to medical imaging thickness, thus improves the performance of TSM on mixed-thickness data (\eg, a mix of thin- and thick-slice CT scans).

Fig.~\ref{fig:operators} illustrates these operators, and Table~\ref{tab:operators} summarizes the computational overhead over no fusion, in terms of parameters and FLOPs. Apart from theoretical FLOPs, we also provide the numeric FLOPs for 3/7-slice inputs to better understand the algorithm complexity in practice. To fairly compare these methods, only FLOPs in 3D backbone are counted, those in 3D-to-2D feature layer and detection heads on 2D feature maps are ignored. Interestingly, additional FLOPs introduced by A3D are marginal given a two-decimal precision.





\begin{table}[tb]
	\caption{Parameters and theoretical (theo.) FLOPs analysis for 3D context fusion operators, in terms of overhead over no fusion, whose parameters and FLOPs are $C_o C_i K^2$ and $\mathcal{O}(DHW C_o C_i K^2)$, respectively.  $D$ denotes the number of slices, $D\times W$ denotes the spatial size of each slice, $C_i$ and $C_o$ denote the input and output channel, and $K$ denotes the kernel size. We also provide the numeric FLOPs of the 3D backbone part for 3/7-slice inputs, \ie, GFLOPs (3/7). Additional FLOPs introduced by A3D are marginal given a two-decimal precision.}\label{tab:operators}
	\centering
	
	\begin{tabular*}{\hsize}{@{}@{\extracolsep{\fill}}lcccccc@{}}
		\toprule
		Operators & 
		No Fusion &
		I3D~\cite{carreira2017quo} &
		P3D~\cite{qiu2017learning} &
		ACS~\cite{yang2021reinventing} & Shift~\cite{lin2019tsm,yang2020alignshift} 
        & A3D (Ours) \\
		\midrule
        Parameters & $1$ & $K$ & $1+C_o/(C_i K)$ & $1$ & $1$ & $1+D^2/(C_o K^2)$\\
        Theo. FLOPs & $1$ & $K$ &$1+C_o/(C_i K)$ &$1$&$1$&$1+D/(C_o K^2)$ \\
        GFLOPs (3) & 40.64 & 78.69 & 67.79 & 40.64 & 40.64 & 40.64 \\
        GFLOPs (7) & 94.83 & 183.61 & 158.18 & 94.83 & 94.83 & 94.83\\
		\bottomrule
	\end{tabular*}

\end{table}

\subsection{Asymmetric 3D Context Fusion (A3D)} \label{sec:a3d}

\begin{algorithm}[tb]
	
	\caption{Asymmetric 3D Context Fusion (A3D)}
	\label{algo:A3D}
	\small
	\KwInput{3D input feature $\boldsymbol{X_i} \in \mathbb{R}^{C_i\times D\times H\times W}$.}
    \KwParameter{asymmetric fusion weight $\boldsymbol{P} \in \mathbb{R}^{D\times D\times C_i}$, \\
    \qquad\qquad\qquad
    2D (pretrained) convolutional kernel $\boldsymbol{W_{2D}} \in \mathbb{R}^{C_i\times C_o\times K\times K}$.}
	\KwOutput{3D output feature $\boldsymbol{X_o} \in \mathbb{R}^{C_o\times D\times H\times W}$.}
	
	\nl $\boldsymbol{W_{3D}}=\textit{unsqueeze}(\boldsymbol{W_{2D}}, \textit{dim}=2) \in \mathbb{R}^{C_i\times C_o\times 1\times K\times K}$,
	
	\nl $\boldsymbol{X}  = \textit{einsum}(``cdhw,dkc\rightarrow ckhw", [\boldsymbol{X_i}, \boldsymbol{P}])\in \mathbb{R}^{C_i\times D\times H\times W}$,
	
	\nl $\boldsymbol{X_o}  = \textit{Conv3D}(\boldsymbol{X},  \textit{kernel}=\boldsymbol{W_{3D}})$.
	
\end{algorithm}

The 3D context fusion operators above are designed to be spatially symmetric, \ie, each 2D slice is transformed identically to ensure these convolution-like operations to be  translation-equivariant. However, in many medical imaging applications, only a few slices are used as model inputs because of memory constraints ($D=3$ or $7$ in this study). In this case, padding (zero or others) on the axial axis induces a significant distribution shift near top and bottom slices. Moreover, convolution-like operations are not truly translation-equivariant~\cite{luo2016understanding} due to limited effective receptive fields. It is 
not necessary to use spatially symmetric operators in pursuit of translation-equivariance for 3D context fusion. 

To address this issue, we propose a novel asymmetric 3D context fusion operator (A3D), which uses different weights to fuse 3D context for each slice. Mathematically, given a 3D input feature $\boldsymbol{X_i} \in \mathbb{R}^{C_i \times D \times H \times W}$, A3D fuses features from different slices by creating dense linear connections within the slice dimension for each channel separately. We introduce a trainable asymmetric fusion weight $P \in \mathbb{R}^{D \times D \times C}$, then
\begin{equation}
    X^{(c)}=P^{(c)}\cdot X_i^{(c)} \in \mathbb{R}^{D \times H \times W}, c\in \{1,...,C\},
\end{equation}
where $P^{(c)} \in \mathbb{R}^{D \times D}$ and $X_i^{(c)} \in \mathbb{R}^{D \times H \times W}$ denotes the channel $c$ of $P$ and $X_i$, respectively, $\cdot$ denotes matrix multiplication. The output $X$ denotes the 3D features after 3D context fusion, it is then transformed by a 3D convolution with $\boldsymbol{W_{3D}}\in \mathbb{R}^{C_i \times C_o\times 1 \times K \times K}$ (as in no fusion). A3D can be implemented using Einstein summation and 3D convolution in lines of code. Einstein summation saves up extra memories occupied by intermediate results of operations such as transposing, therefore makes A3D faster and more memory-efficient. We depict a PyTorch-fashion pseudo-code of A3D in Algorithm \ref{algo:A3D}. Batch dimension is ignored for simplicity, while the algorithm is easily batched by changing $``cdhw,dkc\rightarrow ckhw"$ into $``bcdhw,dkc\rightarrow bckhw"$. A3D is a simple operator that can be plugged into any 3D image model with ease.

To facilitate stable training and faster convergence, the convolution kernels in A3D operation can be initialized with ImageNet~\cite{deng2009imagenet} pretrained weights to take advantage of supervised pretraining. Furthermore, we initialize each channel of asymmetric fusion weight $P^{(c)}$ with a identity matrix $I \in \mathbb{R}^{D \times D}$ added with a random perturbation following uniform distribution in $[-0.1,0.1]$, \ie, the A3D is initialized to be like no fusion before training. 

Compared to symmetric 3D context fusion operators, A3D uses dense linear connections to gather global contextual information along the axial axis (illustrated in Fig.~\ref{fig:operators}~(f)), thus avoids the padding issue around the top and bottom slices. Besides, as depicted in Table~\ref{tab:operators}, A3D introduces negligible computational overhead in terms of both parameters and FLOPs compared with no fusion. Since $D$ is typically much smaller than $C_o$, A3D is more lightweight than I3D~\cite{carreira2017quo} and P3D~\cite{qiu2017learning}. Moreover, as A3D can be implemented with natively supported \textit{einsum}, it is faster than ACS~\cite{yang2021reinventing} and Shift~\cite{lin2019tsm,yang2020alignshift} with channel splitting in actual running time. Note that the A3D is NOT translation-equivariant, as it uses different weights for each output slice to fuse the 3D context from input $D$ slices. However, it significantly outperforms existing symmetric context fusion operators with negligible computational overhead.

\begin{figure}[tb]
    \centering
	\includegraphics[width=0.9\textwidth]{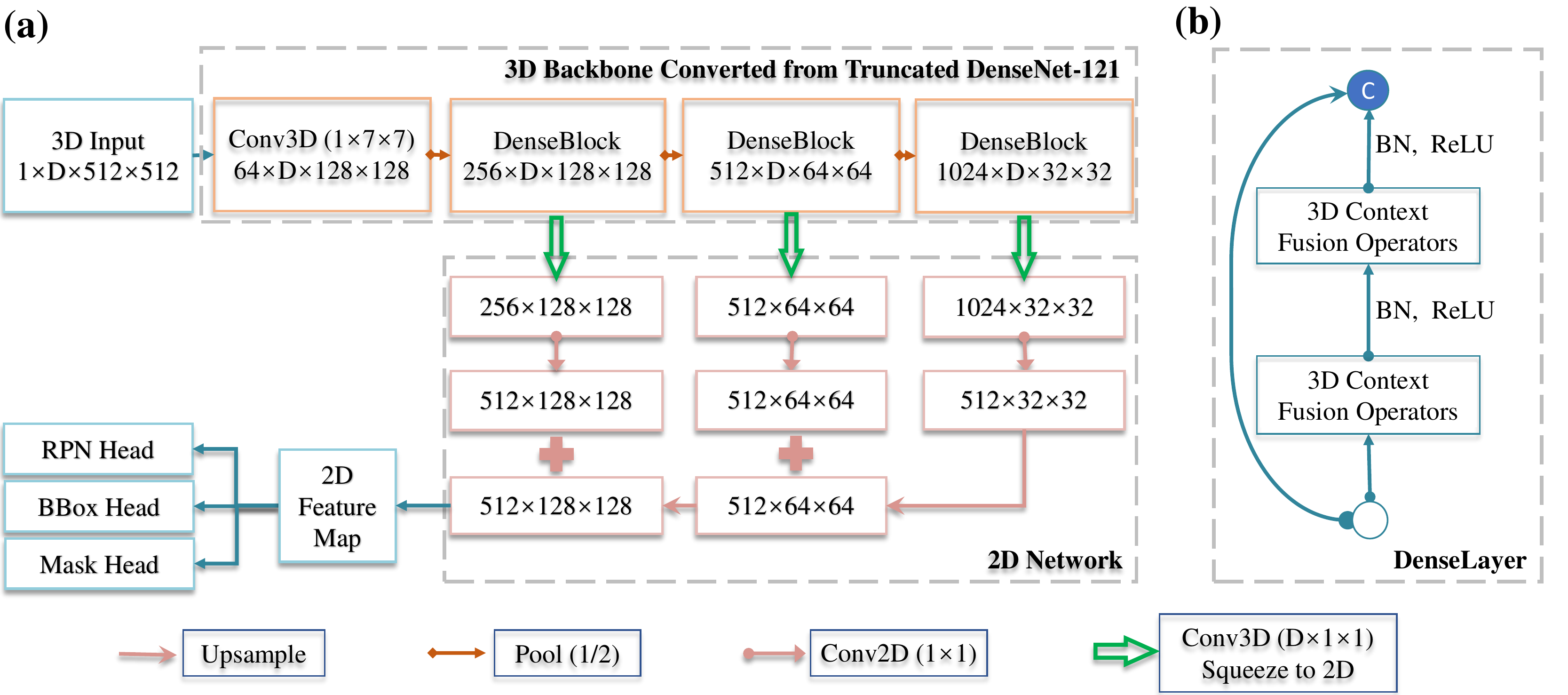}
	\caption{Universal lesion detection model on DeepLesion~\cite{yan2018deep}. The 3D backbone derived from DenseNet-121 \cite{huang2017densely,Yan2019MULANMU} takes a grey-scale 3D input of $D\times 512 \times 512$, where $D$ is the number of slices ($D \in \{3,7\}$ in this study). Features from different scales are collected and fused together in a feature pyramid~\cite{Lin2016FeaturePN}. Detection is based on instance segmentation framework using Mask R-CNN~\cite{He2017MaskR,yang2020alignshift}.} \label{fig:network}
\end{figure}

\subsection{Network Structure for Universal Lesion Detection}

We develop a universal lesion detection model following Mask R-CNN \cite{He2017MaskR}. An overview of our network is depicted in Fig.~\ref{fig:network}. The network consists of a DenseNet-121 \cite{huang2017densely} based 3D backbone with 3D context fusion operators (the proposed A3D or others) plugged in and 2D detection heads. The network backbone takes a gray-scale 3D tensor in shape of $1 \times D \times H \times W$ as input, where $D$ is the number of slices included in each sample ($D \in \{3,7\}$ in this study). Three dense blocks gradually downsample feature maps and increase number of channels while the depth dimension stays at $D$. After spatial and channel-wise unification by upsampling and $D \times 1 \times 1$ convolution, 3D features output by three dense blocks are added together and squeezed to 2D by a $D \times 1 \times 1$ convolution. Finally, the 2D feature maps are used for lesion detection on key slices.

\section{Experiments}
\label{sec:experiments}

\begin{table}[tb]
	
	\caption{Performance evaluated on the large-scale DeepLesion benchmark \cite{yan2018deep} of the proposed A3D versus other 3D context fusion operators, in terms of sensitivities (\%) at various false positives (FPs) per image.}\label{tab:performance-operators}

	\centering
	
	\begin{tabular*}{\hsize}{@{}@{\extracolsep{\fill}}lcccccccc@{}}
		\toprule
		Methods & Slices & 0.5 & 1 & 2 & 4 & 8 & 16 & Avg.[0.5,1,2,4] \\
		\midrule
        No Fusion    	& $\times 3$&72.57 & 79.89 & 86.80 & 91.04 & 94.24 & 96.32 & 82.58\\
        I3D~\cite{carreira2017quo}		& $\times 3$&72.01 & 80.09 & 86.54 & 91.29 & 93.91 & 95.68 & 82.48\\
        P3D~\cite{qiu2017learning}& $\times 3$&62.13&73.21&82.14&88.6&92.37&94.95&76.52\\
       ACS~\cite{yang2021reinventing}& $\times 3$&72.82 & 81.15 & 87.40 & 91.35 & 94.69 & 96.42
        & 83.18\\
        TSM~\cite{lin2019tsm}& $\times 3$&71.80&80.11&86.97&91.10 &93.75&95.56&82.50\\
        AlignShift \cite{yang2020alignshift}	& $\times 3$&73.00 & 81.17 & 87.05 & 91.78 & 94.63 & 95.48 & 83.25\\
		A3D (Ours) & $\times 3$ &\textbf{74.10} &\textbf{81.81} &\textbf{87.87} &\textbf{92.13} &\textbf{94.60}&\textbf{96.50}& \textbf{83.98} \\
		\midrule
        No Fusion	& $\times 7$&73.66 & 82.15 & 87.72 & 91.38 & 93.86 & 95.98 & 83.73\\
        I3D~\cite{carreira2017quo} & $\times 7$&75.37 & 83.43 & 88.68 & 92.20 & 94.52 & 96.07 & 84.92\\
        P3D~\cite{qiu2017learning}& $\times 7$&74.84&82.17&87.57&91.72&94.90&96.23&84.07 \\
        ACS~\cite{yang2021reinventing}& $\times 7$&78.38 & 85.39 & 90.07 & 93.19 & 95.18 & 96.75 & 86.76\\ 
        TSM~\cite{lin2019tsm}& $\times 7$ & 75.98 &83.65 &88.44 &92.14 &94.89 &96.50 &85.05\\
        AlignShift \cite{yang2020alignshift}	& $\times 7$&78.68 & 85.69 & 90.37 & 93.49 & 95.48 & 97.05 & 87.06\\
		A3D (Ours) & $\times 7$ &\textbf{80.27} &\textbf{86.73} &\textbf{91.33} &\textbf{94.12} &\textbf{96.15}&\textbf{97.33}& \textbf{88.11} \\
		\bottomrule
	\end{tabular*}

\end{table}

\subsection{Dataset and Experiment Settings}
DeepLesion dataset \cite{yan2018deep} includes 32,120 axial CT sclies extracted from 10,594 studies of 4,427 patients. There are 32,735 lesions labelled in various organs in total. Each slice contains 1 to 3 lesions, sizes of which range from 0.21 to 342.5mm. RECIST diameter coordinates and bounding boxes are annotated in key slices. Adjacent slices within the range of $\pm$15mm from the key slice are given as contextual information. 

Our experiments are based on the official code of AlignShift~\cite{yang2020alignshift}, and A3D code is merged into the same code repository. Since DeepLesion does not contain pixel-wise segmentation labels, we use GrabCut \cite{rother2004grabcut} to generate weak segmentation labels from RECIST annotations following \cite{yang2020alignshift,zlocha2019improving}. Input CT Hounsfield units are clipped to $[-1024,2050]$ and then normalized to $[-50,205]$. For AlignShift~\cite{yang2020alignshift}, we process the inputs as in its official code since it uses imaging thickness as inputs. For A3D and other counterparts, we normalize the axial thickness of all data to 2mm and resize each slice to $512 \times 512$. In terms of data augmentation, we apply random horizontal flip, shift, rescaling and rotation during the training stage. No test-time augmentation is adopted. We follow the official data split of 70\%/15\%/15\% for training, validation and test, respectively. As per \cite{Yan2019MULANMU,zlocha2019improving,li2019mvp}, the proposed method and its counterparts are evaluated on the test set using sensitivities at various false positive levels (\ie, FROC analysis). We also implement the mentioned 3D context fusion operators to validate the effectiveness of the proposed A3D.

\begin{table}[tb]
	
	\caption{Performance evaluated on the large-scale DeepLesion benchmark \cite{yan2018deep} of the proposed A3D versus previous \emph{state-of-the-art}, in terms of sensitivities (\%) at various false positives (FPs) per image.}\label{tab:deeplesion-sota}

	\centering
	
	\begin{tabular*}{\hsize}{@{}@{\extracolsep{\fill}}lccccccccc@{}}
		\toprule
		Methods & Venue & Slices & 0.5 & 1 & 2 & 4 & 8 & 16 & Avg.[0.5,1,2,4] \\
		\midrule
		3DCE \cite{yan20183d} & ~MICCAI'18 & $\times 27$ & 62.48 & 73.37 &80.70 &85.65 &89.09 &91.06& 75.55\\
		ULDor \cite{tang2019uldor}&~ISBI'19& $\times 1$ & 52.86 & 64.80 & 74.84 & 84.38 & 87.17 & 91.80 & 69.22\\
		V.Attn \cite{wang2019volumetric}&~MICCAI'19&$\times 3$&  69.10 & 77.90 & 83.80 & - & - & - &-\\
		Retina. \cite{zlocha2019improving}&~MICCAI'19&$\times 3$&  72.15 & 80.07 & 86.40 & 90.77 & 94.09 & 96.32 & 82.35\\
		MVP \cite{li2019mvp}&~MICCAI'19 &$\times 3$&  70.01 & 78.77 &84.71 &89.03 &- &- &80.63 \\
		MVP \cite{li2019mvp}&~MICCAI'19 &$\times 9$&  73.83 & 81.82 & 87.60 & 91.30 & - & - & 83.64 \\
		MULAN \cite{Yan2019MULANMU}&~MICCAI'19&$\times 9$&  76.12 & 83.69 & 88.76 & 92.30 & 94.71 & 95.64 & 85.22 \\
		Bou.Maps \cite{li2020bounding}&~MICCAI'20&$\times 3$&  73.32& 81.24&86.75&90.71&-&- & 83.01 \\
		MP3D \cite{zhang2020revisiting}&~MICCAI'20&$\times 9$&  79.60& 85.29& 89.61& 92.45&-&- & 86.74 \\
		AlignShift \cite{yang2020alignshift}&~MICCAI'20 & $\times 3$&73.00 & 81.17 & 87.05 & 91.78 & 94.63 & 95.48 & 83.25\\
		AlignShift \cite{yang2020alignshift}&~MICCAI'20&$\times 7$ & 78.68 & 85.69 & 90.37 & 93.49 & 95.48 & 97.05 & 87.06\\
		ACS \cite{yang2021reinventing}&~JBHI'21 & $\times 3$&72.82 & 81.15 & 87.40 & 91.35 & 94.69 & 96.42 & 83.18\\
		ACS \cite{yang2021reinventing}&~JBHI'21&$\times 7$ &78.38 & 85.39 & 90.07 & 93.19 & 95.18 & 96.75 & 86.76\\ 
		\midrule
		A3D &Ours & $\times 3$ &{74.10} &{81.81} &{87.87} &{92.13} &{94.60}&{96.50}& {83.98} \\
		A3D &Ours & $\times 7$ &\textbf{80.27} &\textbf{86.73} &\textbf{91.33} &\textbf{94.12} &\textbf{96.15}&\textbf{97.33}& \textbf{88.11} \\
		\bottomrule
	\end{tabular*}

\end{table}

\subsection{Performance Analysis}
We compare A3D with a variety of 3D context fusion operators (see Sec.~\ref{sec:preliminary}) on the DeepLesion dataset. Table \ref{tab:performance-operators} gives the detailed performances of A3D and all its counterparts on 3 and 7 slices. A3D delivers superior performances compared with all counterparts on both 3 slices and 7 slices. We attribute this performance boost to A3D's ability of gathering information among globally along the axial axis by creating dense connections among slices, which can be empirically validated by the observation that A3D has a higher performance boost on 7 slices that on 3 slices when compared with the previous \emph{state-of-the-art} AlignShift~\cite{yang2020alignshift} ($+1.05$ vs. $+0.73$) since 7 slices provide more contextual information. Moreover, A3D introduces no padding along the axial axis, this advantage also leads to the performance boost compared to other operators. Note that AlignShift-based model is adaptive to imaging thickness, which is an orthogonal contribution to this study. The asymmetric operation-based methods could be potentially improved by adapting imaging thickness.

Table \ref{tab:deeplesion-sota} shows a performance comparison of A3D and previous \emph{State of the Art}. Without heavy engineering and data augmentations, our proposed method outperforms the previous \emph{state-of-the-art} AlignShift \cite{yang2020alignshift} on both 3 slices and 7 slices by considerable margin. It is worth noting that A3D with image only surpasses MULAN~\cite{Yan2019MULANMU} by nearly 3\% even though it takes less slices and no additional information apart from CT images such as medical report tags and demographic information as inputs.

\begin{figure}[tb]
    \centering
	\includegraphics[width=\linewidth]{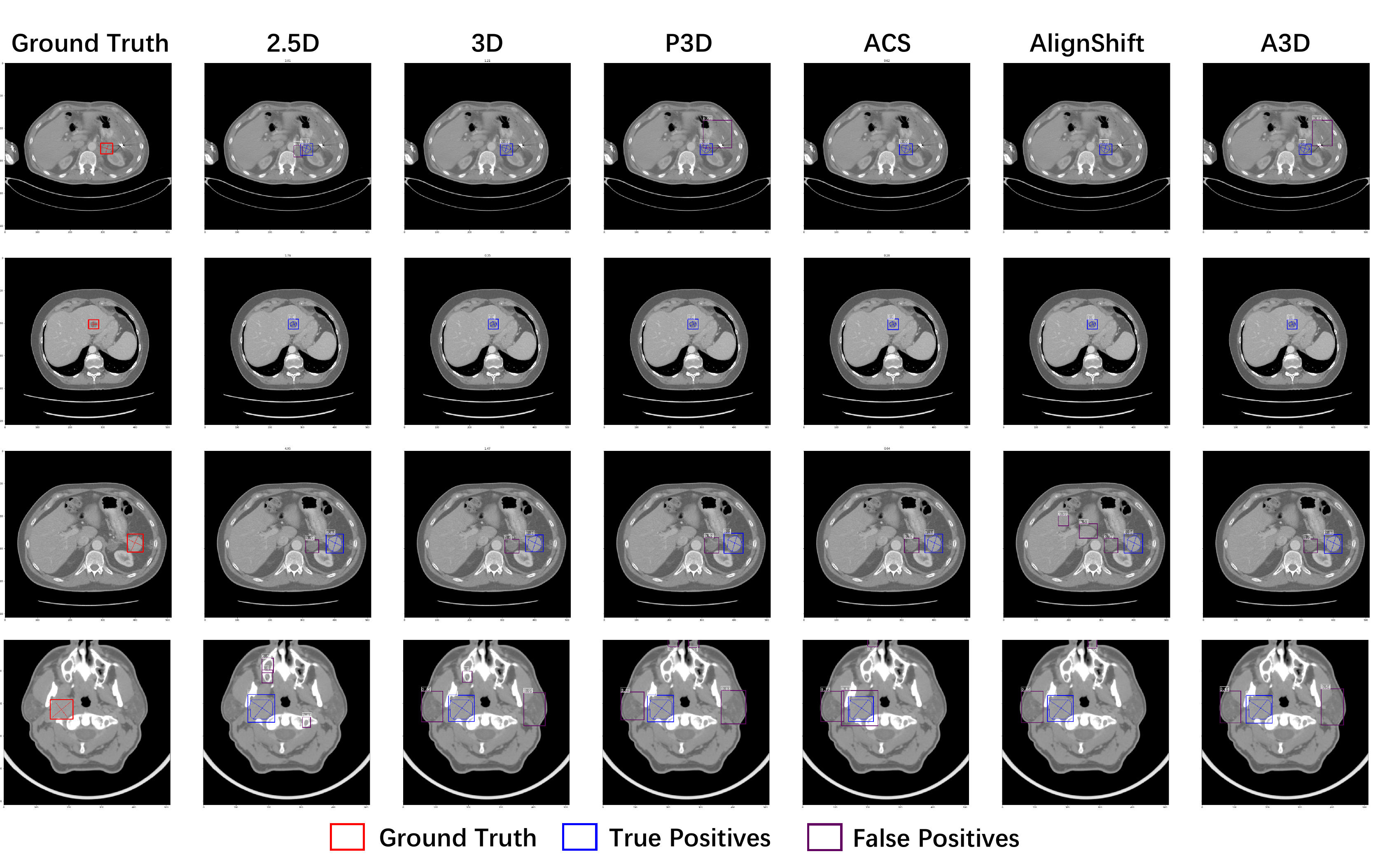}
	\caption{Visualization of DeepLesion slices highlighted with ground truth and predictions generated by different 3D context fusion operators.}
	\label{fig:visualization}
\end{figure}

\section{Conclusion}

In this study, we focus on 3D context fusion operators that enable 2D pretraining, which is an emerging technique that leverages advantages of both 2D pretraining and 3D context modeling. We analyze the unnecessary pursuit of translation-equivariance in existing spatially symmetric 3D context fusion operators especially when only a few 2D slices are used as model inputs. To this end, we further propose a novel asymmetric 3D context fusion operator (A3D) that is translation-equivariant. The A3D significantly outperforms existing symmetric context fusion operators without introducing large computational overhead. Extensive experiments on DeepLesion benchmark validate the effectiveness of the proposed method,  and we establish a new \emph{state of the art} that surpasses prior arts by considerable margins.

\subsubsection{Acknowledgment.}
This work was supported by National Science Foundation of China  (U20B2072, 61976137).

\bibliographystyle{splncs04}
\bibliography{reference}

\begin{thebibliography}{10}
\providecommand{\url}[1]{\texttt{#1}}
\providecommand{\urlprefix}{URL }
\providecommand{\doi}[1]{https://doi.org/#1}

\bibitem{antonelli2021medical}
Antonelli, M., Reinke, A., Bakas, S., et~al.: The medical segmentation
  decathlon. arXiv preprint arXiv:2106.05735  (2021)

\bibitem{carreira2017quo}
Carreira, J., Zisserman, A.: Quo vadis, action recognition? a new model and the
  kinetics dataset. In: CVPR. pp. 6299--6308 (2017)

\bibitem{deng2009imagenet}
Deng, J., Dong, W., Socher, R., Li, L.J., Li, K., Fei-Fei, L.: Imagenet: A
  large-scale hierarchical image database. In: CVPR. pp. 248--255 (2009)

\bibitem{He2017MaskR}
He, K., Gkioxari, G., Doll{\'a}r, P., Girshick, R.B.: Mask r-cnn. ICCV pp.
  2980--2988 (2017)

\bibitem{huang2017densely}
Huang, G., Liu, Z., Van Der~Maaten, L., Weinberger, K.Q.: Densely connected
  convolutional networks. In: CVPR. vol.~1, p.~3 (2017)

\bibitem{jin2020deep}
Jin, L., Yang, J., Kuang, K., Ni, B., Gao, Y., Sun, Y., Gao, P., Ma, W., Tan,
  M., Kang, H., et~al.: Deep-learning-assisted detection and segmentation of
  rib fractures from ct scans: Development and validation of fracnet.
  EBioMedicine  \textbf{62},  103106 (2020)

\bibitem{li2020bounding}
Li, H., Han, H., Zhou, S.K.: Bounding maps for universal lesion detection. In:
  MICCAI. pp. 417--428. Springer (2020)

\bibitem{li2019mvp}
Li, Z., Zhang, S., Zhang, J., Huang, K., Wang, Y., Yu, Y.: Mvp-net: Multi-view
  fpn with position-aware attention for deep universal lesion detection. In:
  MICCAI. pp. 13--21. Springer (2019)

\bibitem{lin2019tsm}
Lin, J., Gan, C., Han, S.: Tsm: Temporal shift module for efficient video
  understanding. In: ICCV. pp. 7083--7093 (2019)

\bibitem{Lin2016FeaturePN}
Lin, T.Y., Doll{\'a}r, P., Girshick, R.B., He, K., Hariharan, B., Belongie,
  S.J.: Feature pyramid networks for object detection. CVPR pp. 936--944 (2016)

\bibitem{litjens2017survey}
Litjens, G., Kooi, T., Bejnordi, B.E., Setio, A.A.A., Ciompi, F., Ghafoorian,
  M., Van Der~Laak, J.A., Van~Ginneken, B., S{\'a}nchez, C.I.: A survey on deep
  learning in medical image analysis. Medical image analysis  \textbf{42},
  60--88 (2017)

\bibitem{luo2016understanding}
Luo, W., Li, Y., Urtasun, R., Zemel, R.S.: Understanding the effective
  receptive field in deep convolutional neural networks. In: NIPS (2016)

\bibitem{qiu2017learning}
Qiu, Z., Yao, T., Mei, T.: Learning spatio-temporal representation with
  pseudo-3d residual networks. In: ICCV. pp. 5533--5541 (2017)

\bibitem{rother2004grabcut}
Rother, C., Kolmogorov, V., Blake, A.: " grabcut" interactive foreground
  extraction using iterated graph cuts. ACM transactions on graphics (TOG)
  \textbf{23}(3),  309--314 (2004)

\bibitem{shen2017deep}
Shen, D., Wu, G., Suk, H.I.: Deep learning in medical image analysis. Annual
  review of biomedical engineering  \textbf{19},  221--248 (2017)

\bibitem{tang2019uldor}
Tang, Y.B., Yan, K., Tang, Y.X., Liu, J., Xiao, J., Summers, R.M.: Uldor: a
  universal lesion detector for ct scans with pseudo masks and hard negative
  example mining. In: ISBI. pp. 833--836. IEEE (2019)

\bibitem{wang2019volumetric}
Wang, X., Han, S., Chen, Y., Gao, D., Vasconcelos, N.: Volumetric attention for
  3d medical image segmentation and detection. In: MICCAI. pp. 175--184.
  Springer (2019)

\bibitem{wei2020mitoem}
Wei, D., Lin, Z., Franco-Barranco, D., Wendt, N., Liu, X., Yin, W., Huang, X.,
  Gupta, A., Jang, W.D., Wang, X., et~al.: Mitoem dataset: Large-scale 3d
  mitochondria instance segmentation from em images. In: MICCAI. pp. 66--76.
  Springer (2020)

\bibitem{yan20183d}
Yan, K., Bagheri, M., Summers, R.M.: 3d context enhanced region-based
  convolutional neural network for end-to-end lesion detection. In: MICCAI. pp.
  511--519. Springer (2018)

\bibitem{Yan2019MULANMU}
Yan, K., Tang, Y., Peng, Y., Sandfort, V., Bagheri, M., Lu, Z., Summers, R.M.:
  Mulan: Multitask universal lesion analysis network for joint lesion
  detection, tagging, and segmentation. In: MICCAI (2019)

\bibitem{yan2018deep}
Yan, K., Wang, X., Lu, L., Zhang, L., Harrison, A.P., Bagheri, M., Summers,
  R.M.: Deep lesion graphs in the wild: relationship learning and organization
  of significant radiology image findings in a diverse large-scale lesion
  database. In: CVPR. pp. 9261--9270 (2018)

\bibitem{yang2020alignshift}
Yang, J., He, Y., Huang, X., Xu, J., Ye, X., Tao, G., Ni, B.: Alignshift:
  bridging the gap of imaging thickness in 3d anisotropic volumes. In: MICCAI.
  pp. 562--572. Springer (2020)

\bibitem{yang2021reinventing}
Yang, J., Huang, X., He, Y., Xu, J., Yang, C., Xu, G., Ni, B.: Reinventing 2d
  convolutions for 3d images. IEEE Journal of Biomedical and Health Informatics
   (2021)

\bibitem{yang2020medmnist}
Yang, J., Shi, R., Ni, B.: Medmnist classification decathlon: A lightweight
  automl benchmark for medical image analysis. In: ISBI (2021)

\bibitem{zhang2020revisiting}
Zhang, S., Xu, J., Chen, Y.C., Ma, J., Li, Z., Wang, Y., Yu, Y.: Revisiting 3d
  context modeling with supervised pre-training for universal lesion detection
  in ct slices. In: MICCAI. pp. 542--551. Springer (2020)

\bibitem{zlocha2019improving}
Zlocha, M., Dou, Q., Glocker, B.: Improving retinanet for ct lesion detection
  with dense masks from weak recist labels. In: MICCAI. pp. 402--410. Springer
  (2019)

\end{thebibliography}

\end{document}